\documentclass[floatfix, preprintnumbers, aps, amsmath, amssymb,    showpacs, 10pt, twocolumn]{revtex4}
\usepackage{graphicx}
\usepackage{dcolumn}
\usepackage{epsfig}
\usepackage{bm}

\begin{document}

\date{\today}

\title{Illuminating the dark ages of the universe: the exact backreaction in the SFRW model and the acceleration of the universe }

\author{Reza Mansouri } \email[On sabbatical leave from Department 
of Physics, Sharif University of Technology, Tehran, and Institute
 for Studies in Physics and Mathematics(IPM), Tehran.]{mansouri@hep.physics.mcgill.ca}

\affiliation{Dept.of Physics, McGill University, Montr\'eal QC,\\
Canada, H3A 2T8}

\begin{abstract}
Within the recently proposed structured FRW model universe the averaged Einstein equations are derived. The backreaction turns out to have an interesting behavior. Its equivalent density and pressure, being proportional, are negative at early times of the dark ages of the universe, and change sign near our present time in our local patch. In addition to explaining the observed dimming of the SNIa it leads to new effects for small cosmic redshifts and also to the difference between the local and global Hubble parameter. Interpreting the backreaction in the FRW-{\it picture}, it is equivalent to a time dependent dark energy with $w = -1$.
\end{abstract}


\maketitle

\newcommand{\eq}[2]{\begin{equation}\label{#1}{#2}\end{equation}}

\par
Take the universe as it is: homogeneous at large scales and inhomogeneous at small scales within structured patches. How are the deviations from the standard homogeneous cosmic fluid in the matter dominated phase of the universe reflected in the cosmological data?  Recent observational data on SNIa imply a larger distance to supernovae than predicted by the conventional FRW universe \cite{r98, p99, sh}, leading to the term acceleration of the universe\cite{r04, f04}, and to the concept of the dark energy.  \\
We have recently proposed\cite{m} the {\it Structured} FRW(SFRW) 
model of the universe as a first step to incorporate the local inhomogeneity of the cosmic fluid into a model universe in accordance with the observational needs. In the SFRW model of the universe the local patches, grown out of the primordial perturbations, and their backreaction on the homogeneous background are modelled exactly as a truncated flat Lomaitre-Tolman-Bondi (LTB) manifold embedded in a FRW universe from which a sphere of the same extent as the LTB patch is removed. As a result of the junction conditions the mean density of any such inhomogeneous patch, with over- and under-dense regions, has to be equal to the density of the FRW bulk\cite{m, km}. Therefore, the Copernican principle is in no way violated and we are led to a model universe where the local patches are distributed homogeneously in the bulk having the same mass as a local FRW patch would have, accounting for all the structures we see grown out of the primordial perturbations within a FRW universe. The analysis of the luminosity distance relation in our structured FRW model showed explicitly a dimming of objects within a patch relative to what it would be inferred from a standard FRW universe\cite{m}. \\    
The local inhomogeneous matter dominated patch has a geometrical domain denoted by $D$ and a hypersurface $\Sigma$ as its boundary to the FRW bulk. Our calculation is based on an exact general relativistic formulation of gluing manifolds.  The inhomogeneous patch containing dust matter is represented by a
flat LTB metric embedded in a pressure-free FRW background universe with the uniform density $\rho_{b}$. We choose the general LTB metric to be written in the synchronous comoving coordinates in the form\cite{km}:
\begin{equation}
ds^2 = -dt^2 + a^2\big[\big(1 + \frac{a' r}{a}\big)^2 \frac{dr^2}{1-k(r)r^2}+
       r^2d\Omega^2 \big],
\end{equation}
where the familiar LTB metric function is now defined as $R(r,t) = a(t,r) .r$.The similarity to the Robertson-Walker metric as now obvious. The overdot and prime denote partial differentiation with respect to $t$ and $r$, respectively, and $k(r)$ is an arbitrary real function such that $k(r)r^2 < +1 $ playing the role of the curvature scalar $k$ in the FRW universe. Hence, the flat LTB is defined by the vanishing of $k(r)$. For a homogeneous universe, $a$ and $k$ don't depend on $r$ and we get the familiar Robertson-Walker metric. In our SFRW universe, the metric outside the inhomogeneous patch, is Robertson-Walker again. The corresponding field equations and the solution for the flat case $k(r) = 0$, can be written in the following familiar form:
\begin{eqnarray}
\big (\frac{\dot a}{a} \big) = {1\over 3} \varrho, \\
\frac{\ddot a}{a} = -\frac{1}{6}\varrho,\\
a(r) = (\frac{3}{4}\varrho)^{1\over 3}(t- t_n(r))^{2\over3}, 
\end{eqnarray}
where we have introduced $\varrho \equiv \frac{6M(r)}{r^3}$. 
The mass $M(r)$ is defined as  
\begin{equation}
M(r) = \int^{R(r,t)}_{0}\rho(r,t)R^{2}dR
     = \frac{1}{6} \overline{\rho}(r,t) R^3, 
\end{equation}
where $\rho(r,t)$ is the density and $\overline {\rho}$, as a function of $r$ and $t$, is an average density up to the radius $R(r,t)$. Note that the volume element in the integral above is not in general equal to the proper volume element of the metric, except for the flat case $k(r) = 0$ we will consider\cite{s}. In contrast, the average density in the patch defined by using the proper volume element will be different from the average above, except for the flat LTB case. The field equations (2-5) are very similar to the familiar Friedmann equations, except for the $r$-dependence of the different quantities. Furthermore, we assume $R'(r,t) = a + ra'>0$ to avoid shell crossing of dust matter during their radial motion. $t_n (r)$ is an arbitrary function of $r$ appearing as an integration 'constant'. This arbitrary function has puzzled different authors who give it the name of 'bang time function' corresponding to the big bang singularity\cite{ce, bo, he}. It has, however, a simple astrophysical meaning within our structured FRW universe. As $R(r,t)$ is playing the role of the radius of our local patch, the time $t = t_n$, leading to $R = 0$, means the time of the onset of the mass condensation or nucleation within the homogeneous cosmic fluid. That is why we have preferred to use the subscript $n$ for it indicating the time of nucleation. As was pointed out in \cite{m}, for a realistic density profile, $t_n$ is a decreasing function of the coordinate $r$ having a maximum at $r = 0$, i.e. at the center of our patch. This means, contrary to the usual interpretation in the literature, that $t > t_n$ for all $0 < r < r_{\Sigma} \equiv L$. Therefore, for all times after the onset of mass condensation within our patch $R(t, r)$ is non-vanishing and for times $t < t_n(r =0)$ we have the full FRW without any structure.  \\
Now, without going into the detailed discussion(see \cite{km, m}), we know already that, assuming there is no thin shell at the boundary of the matching, we must have 
\begin{equation}
\overline{\rho} \stackrel{\Sigma}{=} \rho_b,   
\end{equation}
where $\stackrel{\Sigma}{=}$ means the quantities are to be taken at the boundary to the FRW bulk. We, therefore, are left with the only case imposed by the dynamics of the Einstein equations in which the mean density of a local patch is exactly equal to the density of the background FRW universe: a desired exact dynamical result reflecting the validity of the cosmological 
principle at large, meaning each nucleated patch within the FRW universe have the same average mass density as the bulk. The total mass in a local patch, being equal to the background density times the volume of the patch, is distributed individually due to its self-gravity, leading to overdense structures and voids to compensate it. Assuming again the matter inside each patch to be smoothed out in the form of an inhomogeneous cosmic fluid, we expect it to be overdense at the center, decreasing smoothly to an underdense compensation region, a void, up to the point of matching to the background.\\
The density distribution within a patch must be such that the overdensities of structures are compensated by voids. The nucleation time signals the onset of condensation in the patch which- at least partially- opposes the overall expansion. The running of the function $t_n$ is crucial for the expansion history of the patch and therefore will influence the luminosity of the structures growing within the patch. So far it was shown that $t_n' < 0$\cite{m}. Of course, the nucleation time function is related to the actual mass distribution for which, taking into account the fine structure of the patch including the substructures, we have to rely on the overall observations and the matter power spectrum\cite{zehavi, goodwin, dekel, kocevski}. \\
We envisage now an averaging process in which the inhomogeneities within the local patch are smoothed out and we have again a FRW-type homogeneous modeling of our local patch. The traditional way of doing cosmology is to take the average of the matter distribution in the universe and write down the Einstein equations for it, adding some symmetry requirement. One then solves the equations $G_{\mu \nu} = \langle T_{\mu \nu} \rangle$, assuming homogeneity and isotropy of the mass distribution as the underlying symmetry. This is based on the simplicity principle much used in theoretical physics. As far as the precision of the observations allow, we may go ahead with this simplification. The more exact equation, however, is
$\langle G_{\mu \nu} \rangle = \langle T_{\mu \nu} \rangle$. Calling the difference $G_{\mu \nu} - \langle G_{\mu \nu} \rangle = Q_{\mu \nu}$, one may write the correct equation as $G_{\mu \nu} = \langle T_{\mu \nu} \rangle + Q_{\mu \nu}$. The backreaction term $Q$ has so far been neglected in cosmology because of its smallness. Now that measuring $Q$ is within the range of observational capabilities we have to take it into account. There is, therefore, no need yet to change the underlying general relativity or introduce any mysterious dark energy to mimic $Q$. Of course, the averaging process is neither trivial nor unambiguous, but it is the art of physics to master it. Fortunately, there is an averaging formalism, developed mainly by Thomas Buchert\cite{b, b03, b057, b05}, which can easily be adapted to our LTB patch, having the same mass as the the FRW sphere cut out of it. In this formalism the space-average of any function $f(t,r)$ is defined by
\begin{equation}
\langle f\rangle \equiv {1 \over V_D} \int_D dV f,
\end{equation}
where $dV$ is the proper volume element of the 3-dimensional domain $D$ of the patch we are considering and $V_D$ is its volume. It has been shown\cite{be,b} that in such a mass preserving patch the space-volume average of any function $f(r,t)$ does not commute with its time derivative:
\begin{equation}
\langle f\rangle^{\cdot} - \langle \dot f\rangle = \langle f\theta\rangle -  
\langle f\rangle \langle \theta \rangle,
\end{equation}
where the expansion scalar $\theta$, being equal to the minus of the trace of the second fundamental form of the hypersurface $t = const.$, is now a function of $r$ and $t$. The right hand side trivially vanishes for a FRW universe because of the homogeneity. This fact has far-reaching consequences for observational cosmology in our non-homogeneous neighborhood. The variation of the Hubble function with respect to the red-shift is not so simple any more as in the simple case of FRW universe\cite{m}. This affects a lot of observational data processing which so far has been done assuming homogeneity of the universe. Depending on the smoothing width $\Delta z$, the bins, and the matter power spectrum there may be huge effects due to the non-commutativity of the averaging process\cite{e}.\\
The averaged scale factor is defined using the volume of our patch $D$ by $a_D \equiv V(t)_D^{1 \over 3}$. Now it can be shown that\cite{b, be}
\begin{equation}
\theta_D \equiv \langle \theta\rangle \equiv  {\dot V \over V} = 
           3{\dot a_D \over a_D} = 3 H_D.
\end{equation}
where we have used the notation $\dot a_D \equiv {d\over dt}a_D$, and denoted the average Hubble function as $H_D$. Averaging over the local patch means we are taking it as an effective FRW patch. Therefore all the derived quantities should be based on the average value $a_D$. This is why we take the above definition for the mean Hubble parameter and not $\langle \frac{\dot a}{a} \rangle$, which is different from $\frac{{\dot a}_D}{a_D}$. A similar difference holds for the second derivative of $a$: 
\begin{equation} 
 \langle\frac{\ddot a}{a} \rangle \not = \frac{\langle \ddot a \rangle} 
{\langle a \rangle} \not = {\ddot a_D \over a_D}.
\end{equation}
Therefore, the definition of the averaged deceleration parameter is not without ambiguity, specially because there is no nice relation like (9) for the deceleration parameter. To choose the most appropriate definition, we make recourse to the fact that in the averaging process we are taking our patch to be homogeneous and FRW-like. Therefore, in averaging the redshift as a function $a$, we always encounter $a_D$ and its time derivatives $\dot a_D$ and $\ddot a_D$. This justifies the above definition of the mean Hubble parameter and motivates us to make the following definition for the deceleration parameter:  
\begin{equation}
q_D  = - \frac{{\ddot a}_D a_D}{{\dot a_D}^2} =  - \frac{{\ddot a}_D}{a_D}\frac{1}{H_D^2},
\end{equation}
as was done in the literature so far\cite{b, hs,nt,kmr,s}. Now, we are ready to take the average of the Einstein equations in our local patch to see how the mean field equations will look like and what are the differences to the simple FRW field equations. The emergence of a crucial term in the mean Einstein equations, the so called {\it backreaction} term, is interesting. Buchert's backreaction term is defined by\cite{b, b05} 
\begin{eqnarray}
Q = \langle \sigma^2 \rangle - {1\over3}\langle (\theta - \langle \theta \rangle)^2 \rangle \\
 = \langle \sigma^2 \rangle - {1\over3}[\langle \theta^2\rangle - \theta_D^2],
\end{eqnarray}
where $\sigma$ is the shear scalar and $\theta$ is the expansion. 
Although $\theta_D$ and $H_D$ are proportional, $\langle \theta^2 \rangle$ and $\langle H^2 \rangle$ are not. Hence, the relations (13, 14) can not be written in terms of $H$, as was done in\cite{nt}. The averages of the Einstein equations using the Hamiltonian constraint and the Raychaudhuri equation, taking into account the subtleties of the observation just mentioned, is then written in the following form\cite{b, b05}:
\begin{eqnarray}
\big({\dot a_D \over a_D} \big)^2 = {1\over3} (\rho_b + Q)  \\
\frac{\ddot a_D}{a_D} = -{1\over6}(\rho_b + 4Q),
\end{eqnarray}
where we have set $\langle \rho \rangle = \rho_b$, the density of the background FRW universe, as a result of the junction conditions reflected in the eq.(6). Note that in the so-called Friedmann equation (15) the local Hubble parameter enters instead of the global one $H_b$. The effect of the backreaction within the local patch is realized as an effective perfect fluid with the equation of state
\begin{eqnarray}
\rho_Q = {4\over3} p_Q. 
\end{eqnarray}
The backreaction term $Q$ can not yet be considered as representing dark energy in the FRW-{\it picture}. The $\rho_b$ appearing in the field equation of our SFRW model is the total background energy density, i.e. $\rho_b = \rho_M + \Lambda$, where we have chosen $\rho_M$ for the matter density. 
The curvature term is set equal zero to have a simple flat universe. But let us first investigate the sign of $Q$ which is crucial for the interpretation of these averaged equations. As the running of the density and the nucleation time $t_n$ influence the mean values of the Hubble parameter and the shear scalar, the sign of $Q$ is determined by the balance between the mean values of the shear and the term related to the mean values of the Hubble parameter and the expansion scalar in a complex manner depending of the running of the density and the nucleation time. Given this complex behavior of the backreaction term, let us approximate $t_n$ in the following way:
\begin{equation}
t_n = t_0 -{\beta \over 2} r^2.
\end{equation}
For $\beta > 0$ the above expansion satisfies all the necessary conditions to be fullfilled by $t_n$ within the SFRW model universe\cite{m}. It happens that this behavior corresponds to the special parabolic case of \cite{s} formulated in their step 5 and
illustrated in their figure 7.a. The calculation of different terms in $Q$
is messy but exactly doable. The result is
\begin{equation}
Q = {6\over L^3}\frac{A}{B} + {4\over 3}\frac{C}{B^2},
\end{equation}
where
\begin{equation} \nonumber
A = -5.13L^3 +9.12{2t\over \beta}L - 0.95({2t\over \beta})^{3\over 2} \arctan(\sqrt
{\beta \over 2t} L)  
\end{equation}
 \begin{equation}\nonumber
 + 0.08 ({2t \over \beta})^{3\over 2} \arctan \sqrt {7\over 3}(\sqrt {\beta\over 2t} L),
\end{equation}
\begin{equation}
\nonumber B = 2t^2 + 2 \beta tL^2 +{1\over 2} \beta^2 L^4,\hspace{1 cm}
 C = (2t + \beta L^2)^2.
\end{equation} 
To understand its behavior, we determine its sign for two limiting cases: at the onset of nucleation, i.e. $t - t_0 \ll \beta L^2$, where the coordinate dependence of different quantities has the biggest effect, and at the present time $t - t_0 \gg \beta L^2$. From the exact result of $Q$ we obtain
\begin{eqnarray}
t - t_0 \ll \beta L^2:  Q \simeq -50.26 \frac{1}{\beta^2 L^4} < 0, \\
t - t_0 \gg \beta L^2:  Q \simeq +49.75 \frac{1}{\beta L^2}\frac{1}{t} > 0.
\end{eqnarray}
 We, therefore, conclude that the backreaction has its strongest
effect on the onset of mass condensation at the beginning of the cosmic dark ages and after the density contrast increases to a proper value, where its effective density and pressure are negative. It then changes somewhere, probably at the end of the dark ages, the sign and behaves as a normal fluid.\\
Negative values of $Q$ at the early stages of mass condensation is a novel effect, even though it is just for a local patch. It has the effect of 
reducing the Hubble parameter and producing a negative pressure for a range of expansion time with the effect of dimming of the cosmic objects in our vicinity. Sometime near our present cosmic time the effect reverses and has to lead to new effects for small redshifts, which are somehow opposed to the acceleration of the universe. Remember that these effects are just within our local patch and in interpreting data along the light cone one must be cautious. The mere fact that local Hubble parameter, $H_D$, may be less or greater than the global one, depending on the behavior of $Q$, is interesting and should be taken into account in announcing the $H$ values. \\
A comparison to the dark energy concept is possible if we switch from the SFRW-picture to the FRW-picture. Let us denote $ -Q = \Lambda$, where $\Lambda$ is now a function of time and space being most of the time positive. Then we have the following equations at our disposal
\begin{eqnarray}
H_D^2 = \big({\dot a_D \over a_D} \big)^2 = {1\over3} \rho_M  \\
H_b^2 = \big({\dot a \over a} \big)^2 = {1\over3} (\rho_M + \Lambda)  \\
\frac{\ddot a_D}{a_D} = -{1\over6}(\rho_M - 3\Lambda).
\end{eqnarray}
Observational cosmologist, using the FRW-picture are used to $H_b$ equation, but taking both the values of $H_b$ and $\rho_b = \rho_M + \Lambda$ from observational data. For the interpretation of the dimming of cosmic objects, however, there is no other way than to use the third equation above. In this picture our backreaction can be interpreted as a time  dependent cosmological constant having $w = -1$. The actual SFRW picture, however, gives us a much wider spectrum of information we should be aware of. In the same picture we may say that the backreaction produces a dissipative pressure or anti-frictional force $Q = -\Lambda$ along the line of reasoning in\cite{ds}. 
A realistic structured FRW universe not only explains the dimming of cosmological objects but also leads to new effects which should be looked for in the huge data already existent. \\

 I would like to thank Thomas Buchert for indicating the non-vanishing
character of the shear in an early draft of the paper. It is also a pleasure to thank members of the cosmology group at McGill University, specially Robert Brandenberger and McGill Physics department for the hospitality.


\begin{thebibliography}{}
\bibitem{r98} A. G. Riess et. al., Astron. J., 116, 10-09 (1998); astro-ph/ 9805201
\bibitem{p99} S. Perlmutter et. al., Astrophys.J., 517, 565 (1999); 
                        astro-ph/981.
\bibitem{sh} Charles A. Shapiro and Michel S. Turner, astro-ph/0512586.
\bibitem{r04} A.G. Riess et. al., Astrophys. J., 607, 665 (2004).
\bibitem{f04} A. V. Fillipenko, astro-ph/0410609.
\bibitem{m} Reza Mansouri, astro-ph/0512605.
\bibitem{km} S. Khakshournia and R. Mansouri, Phys. Rev. D65, 027302, 2003;
\bibitem{s} R. A. Sussman and Luis Garcia Trujillo, Class.Quant.Grav. 19, 2897, 2001; gr-qc/0105081.

\bibitem{ce} M. Celerier, A\&A, 353, 63, 2000; astro-ph/0512103.
\bibitem{bo} K. Bolejko, astro-ph/0512103.
\bibitem{he} C. Hellaby and A. Krasinsky, gr-qc/0510093 (to be published in Phys. Rev. D).
             gr-qc/0307023. 
\bibitem{zehavi} I. Zehavi et. al. Astrophys. J. 503, 483, 1998; astro-ph/9802252.
\bibitem{goodwin} S.P. Goodwin, P. A. Thomas, A. J. Barber, J. Gibbon, and L. I. Onuora, astro-ph/9906187.
\bibitem{dekel} A. Dekel, ARA\&A, 32, 371, 1994.
\bibitem{kocevski} D. D. Kocevski, H. Ebeling, Ch. R. Mullis, and R. Brent                      Tully, astro -ph/0512321.
\bibitem{b} T. Buchert, Gen. Rel. Grav. 32, 105, 2000 and 33, 1381, 2001;                  gr-qc/9906015 and gr-qc/0102049;                                 
\bibitem{b03} T. Buchert, Phys. Rev. Lett. 90, 031101, 2003; gr-qc/0210045; 
\bibitem{b057} T. Buchert, Class. Quant. Grav. 22, L113, 2005; gr-qc/0507028; 
\bibitem{b05} gr-qc/0509124.
\bibitem{be} Thomas Buchert and Juergen Ehlers, A\&A, 320, 1, 1997. 
\bibitem{e}D. J. Eisenstein et.al., Astrophys.J. 633, 560, 2005; astro-ph/0501171.
\bibitem{kmr} E. W. Kolb, S. Matarrese, and A. Riotto, astro-ph/0506534.
\bibitem{hs} Christopher M. Hirata and Uros Seljak, astro-ph/0503582.
\bibitem{nt} Y. Nambu and M. Tanimoto, gr-qc/0507057.
\bibitem{ds} Dominik J. Schwarz, astro-ph/0209584 (talk given at CERN-TH-2002-246).
\end{thebibliography}
\end{document}